\documentclass{iopart}
\expandafter\let\csname equation*\endcsname\relax
\expandafter\let\csname endequation*\endcsname\relax
\usepackage{euscript,amsmath,amssymb,amsfonts,graphicx,bm}
\usepackage{subcaption}
  \usepackage{paralist}
  \usepackage{epstopdf}
  \usepackage{graphics} 
 \usepackage[colorlinks=true]{hyperref}
 \hypersetup{urlcolor=blue, citecolor=red}
 \usepackage[latin1]{inputenc}

\eqnobysec

\renewcommand{\e}{{\rm e}}

\renewcommand{\P}{\mathbb{P}}
\newcommand{\R}{\mathbb{R}}

\newcommand{\E}{\mathbb{E}}

\renewcommand{\u}{\mathbf{u}}
\newcommand{\U}{\mathbf{U}}
\newcommand{\calU}{\mathcal{U}}

\def\calT{\mathcal{T}}
\def\P{\mathbb{P}}

\newcommand{\x}{\mathbf{x}}
\newcommand{\X}{\mathbf{X}}

\newcommand{\Markov}[2]{\underset{#1}{\overset{#2}{\rightleftharpoons}}}

\begin{document}

\title[Global resetting and emergent correlations]{Global resetting and emergent correlations: exit statistics in an interval }

\author{Paul C. Bressloff}
\address{Department of Mathematics, Imperial College London, London SW7 2AZ, UK.}
\ead{p.bressloff@imperial.ac.uk}

\date{\today}

\begin{abstract}
There is considerable current interest in the emergence of statistical correlations within a population of otherwise non-interacting Brownian particles subject to a common fluctuating environment or drive. Examples include global stochastic resetting, switching confining potentials, fluctuating diffusivities, and stochastically gated boundaries. Most studies have focused on the analytical structure of the stationary joint probability density (assuming it exists). In this paper, we extend previous work on the exit statistics of multiple particles in stochastically gated domains to the case of global resetting in an interval with absorbing boundaries at both ends. First, we use a generalised It\^o's lemma to derive a hierarchy of boundary value problems (BVPs) for the joint splitting probability that all particles exit from the same end of the interval. The BVPs form a nested sequence with respect to the initial number of particles $M$. We explicitly solve the BVP for a pair of particles ($M=2$) and use this to illustrate the emergence of pairwise correlations. Second, we show how the BVP for the splitting probability of $M$ Brownian particles can be mapped onto the $M$th order moment equation of a stochastic diffusion equation with resetting. We thus establish a general mathematical framework to study exit problems for globally-driven particle systems.
\end{abstract}

\maketitle

\section{Introduction}

\begin{figure}[t!]
\centering
\includegraphics[width=14cm]{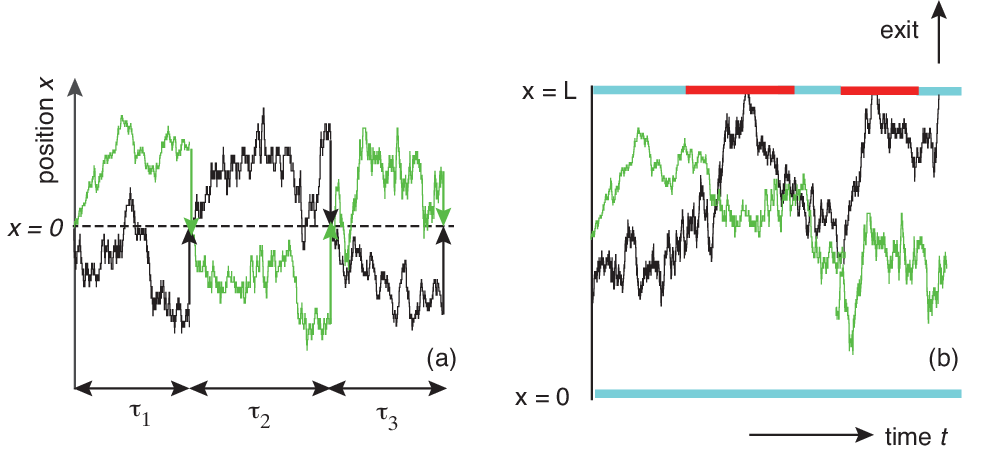} 
\caption{Two examples of globally driven Brownian motion. (a) Pair of Brownian particles diffusing in $\R$ and subject to global resetting. At a random sequence of times $T_n$ the positions of both particles are reset to the origin. The inter-event times $\tau_n=T_{n+1}-T_n$ are exponentially distributed. (b) Pair of Brownian particles in the interval $[0,L]$ with an open boundary at $x=0$ and a stochastically gated boundary at $x=L$. The latter switches between an open state (blue) and a closed state (red) according to a two-state Markov chain. Also shown is the exit of one particle through the open gate.}
\label{fig1}
\end{figure}

Recently, there has been considerable interest in so-called dynamically emergent correlations (DEC), in which a population of otherwise non-interacting Brownian particles develop statistical correlations due to a common external environment or drive that
fluctuates stochastically and independently of the particles, see the recent review \cite{Majumdar26} and references therein. Two major examples are global resetting \cite{Biroli23,Mauro26,Biroli26a,Boyer26}
and a switching harmonic potential \cite{Biroli24b,Sabha24,Biroli26b}. In the former case, all particles are simultaneously returned to a common fixed initial position at a random sequence of times that is typically generated by a Poisson process, see Fig. \ref{fig1}(a), whereas in the latter case all particles are subjected to a common confining potential that randomly switches between two or more distinct configurations. A third mechanism is to take the diffusivity of all the particles to be a globally fluctuating function of time \cite{Mesquita25a}. One particular focus of DEC is to characterise various macroscopic and microscopic observables associated with the non-equilibrium stationary state (NESS) corresponding to the long time limit of the joint probability density function. These include the average density profile of the Brownian gas, the distribution of the position of the $k$th rightmost particle, and the spacing distribution between
two successive particles. In many cases, one finds that there is an underlying analytical structure that expresses the NESS as the weighted integral over products of conditional single particle probability densities.

The observation that a fluctuating common environment can induce statistical correlations in a non-interacting Brownian gas was also made several years ago within the context of diffusion in stochastically gated domains \cite{Bressloff15a,Bressloff15b,Lawley16,Bressloff16,Bressloff16a}. Consider, for example, $M$ particles diffusing in the interval $[0,L]$ with an open boundary at $x=0$ and a stochastically gated boundary at $x=L$, see Fig. \ref{fig1}(b). Let $J(t)\in \{0,1\}$ denote the state of the gate at time $t$ with $J(t)$ evolving according to a two-state Markov chain with switching rates $\alpha,\beta$:
\begin{equation}
\label{M}
0\Markov{\alpha}{\beta}1.
\end{equation}
Assume that the boundary at $x=L$ is totally absorbing (open) when $J(t)=0$ and totally reflecting (closed) when $J(t)=1$. In this system, the commonly fluctuating environment is the stochastic gate. However, in contrast to recent examples of DEC, there does not exist a non-trivial NESS, since all of the particles are eventually absorbed with probability one. Nevertheless, emergent correlations can be studied by considering the joint exit statistics. One particular quantity of interest is the splitting probability $\pi^{(M)}(x_{0,1},\ldots x_{0,M})$ that all $M$ particles are absorbed at the gated end $x=L$ given the initial positions $X_m(0)=x_{0,m}$, $m=1,\ldots,M$ \cite{Bressloff15a,Lawley16}. Using probabilistic arguments based on a generalised It\^o's lemma, it can be proven that $\pi^{(M)}$ satisfies a non-trivial boundary value problem (BVP) in which $\pi^{(M)}$ couples to $\pi^{(M-1)}$ at $x=L$, where $\pi^{(M-1)}$ is the corresponding splitting probability for an initial set of $M-1$ particles. Moreover, the resulting solution cannot be decomposed into a product of single-particle splitting probabilities:
\begin{equation}
\pi^{(M)}(\x_0)\neq \prod_{m=1}^M\pi(x_{0,m}),
\end{equation}
where $\pi_j(x_0)$ is the splitting probability that a single independent particle starting at $x_0$ exits at $x=L$. 

In this paper, we show that an analogous mathematical structure arises in the case of $M$ Brownian particles with global resetting and totally absorbing boundaries at both ends. We begin by revisiting the well-studied example of a single Brownian particle with stochastic resetting in an interval \cite{Pal19}, see section 2. We formulate the problem in a form that is then extended to $M$ particles in section 3. We use a generalised It\^o's lemma and proof by induction to derive the BVP for the splitting probability $\pi^{(M)} (\x_0)$ that all $M$ particles exit at the end $x=L$ given the initial positions $\X(0)=\x_0$ and reset positions $\x_r$. We explicitly calculate $\pi^{(2)}$ for a pair of particles ($M=2$) in section 4, and use this to illustrate the effects of statistical correlations. In section 5 we show how the BVP for $\pi^{(M)}$ can be mapped onto an $M$th-order moment equation for a stochastic partial differential equation (SPDE) consisting of the diffusion equation subject to stochastic resetting. (An analogous relationship between probabilistic models and SPDEs holds for stochastically gated diffusion \cite{Bressloff15a,Lawley16}.) Finally, in appendix A we collect together previous results concerning stochastically gated diffusion in order to highlight the general nature of the mathematical approach used in this paper to analyse dynamically emergent correlations in exit statistics.

\section{Single Brownian particle in an interval with resetting}

Consider a single Brownian particle in the interval $[0,L]$ with totally absorbing boundaries at both ends.  Let $X(t)\in [0,L]$ denote the position of the particle at time $t$ (assuming that the particle hasn't yet exited the domain). In addition, suppose that $X(t)$ instantaneously resets to a fixed reset position $x_r$ at a constant rate $r$. Take $T_{n}$ to be the $n$th resetting time with $n\geq 1$. The inter-reset intervals $\tau_{n}=T_{n+1}-T_{n}$ are exponentially distributed random variables with 
\begin{equation}
\P[\tau_{n}\in [\tau,\tau+d\tau]]=r\e^{-r\tau}d\tau.
\end{equation}
Following Refs. \cite{Magdziarz22,Bressloff24a} we express the evolution of $X(t)$ as the jump-diffusion process
\begin{eqnarray}
dX(t)&=\sqrt{2D}d{W}(t) + (x_{r}-X(t^-))dN(t),
\label{dX}
\end{eqnarray}
where $D$ is the diffusivity, $W(t)$ is a Wiener process with
\begin{eqnarray}
\langle W(t)\rangle = 0,\quad \langle W(t)W(s)\rangle =\min\{t,s\},
\end{eqnarray}
and
\begin{equation}
\label{dNt}
dN(t):=h(t)dt=\sum_{n\geq 1}\delta(t-{T}_n)dt.
\end{equation}
Note that $N(t)$ is a Poisson process with rate $r$ such that if $N(T_n^-)=n-1$ then $N(T_n)=n$. This implies that the solution $X(t)$ is also right-continuous with $X(T_n)=x_r$. In standard formulations of stochastic resetting at the single-particle level \cite{Evans20}, the stochastic differential equation (SDE) (\ref{dX}) is averaged over multiple realisations of the Poisson resetting process. Since $\P[{T}_{n}\in [t,t+dt]]=rdt$, equation (\ref{dX}) becomes
\begin{subequations}
\label{SDEres3}
\begin{eqnarray}
dX(t)&=&\sqrt{2D}d{W}  
\quad  \mbox{ with probability } 1-rdt, \\
dX(t)&=&x_{r}-X(t) \mbox{ with probability } rdt.\end{eqnarray}
\end{subequations}
One of the useful features of the SDE (\ref{dX}) is that one can write down a generalised It\^o's lemma \cite{Bressloff24a}. Let $f(x,t)$ be an arbitrary smooth bounded test function on $[0,L]\times [0,\infty)$. Away from the boundaries, the infinitesimal $df(X(t),t)$ can be decomposed as
\begin{eqnarray}
\fl df(X(t),t)&=&\bigg [\partial_t f(X(t),t)+D\partial^2_{x}f(X(t),t)\bigg ]dt+\sqrt{2D}\partial_x f(X(t),t)dW(t)\nonumber \\
\fl &&\quad +\bigg [f(x_r,t)-f(X(t^-),t^-)\bigg ]dN(t).
\label{lemma}
\end{eqnarray}

Let $p(x,t|x_0)$ be the probability density for the particle to be at position $X(t)=x$ at time $t$ given the initial position $X(0)=x_0$ (and the reset position $x_r$). The corresponding survival probability is
\begin{equation}
Q(x_0,t)=\int_0^L p(x,t|x_0)dx.
\end{equation}
(For ease of notation, we drop the explicit dependence of $p$ and $Q$ on the reset position $x_r$.) 
The density satisfies the forward Kolmogorov equation \cite{Evans11a,Evans11b}
\begin{eqnarray}
\label{FPreset}
\fl \frac{\partial p(x,t|x_0)}{\partial t}=D\frac{\partial^2 p(x,t|x_0)}{\partial x^2} +
r[\delta(x-x_r)Q(x_0,t)-p(x,t|x_0)]  ,
\end{eqnarray}
together with the initial condition $p(x,0|x_0)=\delta(x-x_0)$ and the Dirichlet boundary conditions
\begin{equation}
p(0,t|x_0)=0=p(L,t|x_0).
\end{equation}
One way to derive equation (\ref{FPreset}) is to apply the generalised It\^o's lemma to the identity
\begin{eqnarray}
p(x,t|x_0)=\E[\delta(x-X(t)){\bf 1}_{\calT>t}|X(0)=x_0 ],
\end{eqnarray}
where $\calT$ is the FPT
\begin{equation}
\calT=\inf\{t>0, X(t)\in \{0,L\}\}
\end{equation}
and the indicator function ${\bf 1}_{\calT>t}$ ensures that the particle hasn't exited up to time $t$. Here $\E[\cdot]$ denotes the joint expectation with respect to the Gaussian white noise and resetting processes \cite{Bressloff24a}. The corresponding backward Kolmogorov equation is \cite{Evans11b}
\begin{eqnarray}
\label{BFPreset}
 \frac{\partial p(x,t|x_0)}{\partial t}=D\frac{\partial^2 p(x,t|x_0)}{\partial x_0^2} +
r[p(x,t|x_r)-p(x,t|x_0)]  ,
\end{eqnarray}
with $p(x,t|0)=p(x,t|L)=0$.
Equation (\ref{BFPreset}) can either be derived from the forward equation (\ref{FPreset}) using the Chapman-Kolmogorov integral formula or by another application of the generalised It\^o's lemma.

In Ref. \cite{Pal19}, the FPT problem for exit at the ends $x=0,L$ was analysed extensively. Here we consider an alternative formulation that will be particularly useful when considering correlations in the exit statistics for more than one particle with global resetting. Let $\pi(x_0;x_r)$ be the splitting probability that the particle starting at $x_0$ and resetting to $x_r$ is ultimately absorbed at the end $x=L$ rather than $x=0$.. (It will be convenient in the subsequent analysis to keep track of how the splitting probability depends on $x_r$.) We can relate $\pi(x_0;x_r)$ to the probability density $p(x,t|x_0)$ according to
\begin{equation}
\pi(x_0;x_r)=-D\int_0^{\infty}\frac{\partial p(L,t|x_0)}{\partial x}dt.
\label{defpi}
\end{equation}
The right-hand side is the total  time-integrated probability flux through the end $x=L$. 
Introducing the Laplace transform
\begin{equation}
\widetilde{p}(x,s|x_0)=\int_0^{\infty} \e^{-st}p(x,t|x_0)dt,
\end{equation}
we see that
\begin{equation}
\pi(x_0;x_r)=-D\lim_{s\rightarrow 0} \frac{\partial \widetilde{p}(L,s|x_0)}{\partial x}.
\end{equation}
Hence, one way to determine the splitting probability is to solve the forward Kolmogorov equation (\ref{FPreset}) in Laplace space. This is the approach taken in Ref. \cite{Pal19}. An alternative method is to use the backward Kolmogorov equation. Differentiating both sides of (\ref{BFPreset}) with respect to time $x$, setting $x=L$ and then integrating with respect to $t$ yields the boundary value problem
\begin{subequations}
\begin{equation}
\label{eqpia}
D\frac{d^2\pi(x_0;x_r)}{dx_0^2}+r[\pi(x_r)-\pi(x_0;x_r)]=0,
\end{equation}
with $\pi(x_r)=\pi(x_0;x_r)_{x_0=x_r}$,
supplemented by the boundary conditions
\begin{equation}
\label{eqpib}
\pi(0;x_r)=0,\quad \pi(L;x_r)=1.
\end{equation}
\end{subequations}
The solution is given by
\begin{equation}
\fl \pi(x_0;x_r)=\pi(x_r)(1-\cosh \xi x_0)]+\frac{1-\pi(x_r)(1-\cosh \xi L)}{\sinh\xi L}\sinh \xi x_0,
\label{spi1}
\end{equation}
where $\xi=\sqrt{r/D}$. Finally, setting $x_0=x_r$ we obtain a self-consistency condition for $\pi(x_r)$ which can be solved to give
\begin{equation}
\pi(x_r)=\frac{\sinh \xi x_r}{\sinh \xi x_r+\sinh \xi(L-x_r)}.
\label{spi2}
\end{equation}
We thus recover the corresponding result derived in Ref. \cite{Pal19}.

In anticipation of the multi-particle case, it is useful to derive the following result, which is a version of a Feynman-Kac formula:
\begin{equation}
\pi(x_0;x_r)=\E[\pi(X(t);x_r)],
\label{lemma2}
\end{equation}
under the constraint $X(0)=x_0$.
We proceed using the generalised Ito's lemma (\ref{lemma}). That is,
\begin{eqnarray}
d\pi(X(t);x_r)&=&\bigg [D\partial^2_{x}\pi(X(t);x_r)\bigg ]dt+\sqrt{2D}\partial_x \pi(X(t);x_r)dW(t)\nonumber \\
 &&\quad +\bigg [\pi(x_r)-\pi(X(t^-);x_r)\bigg ]dN(t).
\label{pilemma}
\end{eqnarray}
Since the stochastic process is non-anticipatory, we can take expectations of both sides using $\E[dW(t)]=0$ and $\E[dN(t)]=r$. This gives
\begin{eqnarray}
\fl \frac{d\E[\pi(X(t);x_r)]}{dt}&=& \E\bigg [D\partial^2_{x}\pi(X(t);x_r) +r [\pi(x_r)-\pi(X(t);x_r)\bigg ] .
\end{eqnarray}
Equation (\ref{eqpia}) implies that the right-hand side vanishes. Hence, $\E[\pi(X(t);x_r)]$ is independent of time $t$, $t\geq 0$, and equation (\ref{lemma2}) follows. A further result is obtained by noting that the stochastic process satisfies the strong Markov property\footnote{The strong Markov property is an extension of the Markov property, whereby a stochastic process's future behaviour, conditioned on its present state at a random stopping time $\tau$, is independent of its past.}. In particular, equation (\ref{lemma2}) still holds when $t$ is taken to be the FPT $\tau_1$ that the particle is absorbed at either end. (More precisely, we consider $\min\{t,\tau_1\}$ and then take the limit $t\rightarrow \infty$.) The boundary conditions (\ref{eqpib}) imply that $\pi(X(\tau_1);x_r)=1$ for $X(\tau_1)=L$ and $\pi(X(\tau_1);x_r)=0$ for $X(\tau_1)=0$. We deduce that the solution of the BVP (\ref{eqpia}) and (\ref{eqpib}) is precisely the splitting probability for exit at $x=L$:
\begin{equation}
\pi(x_0;x_r)=\P[\mbox{particle exits at }x=L|X(0)=x_0].
\end{equation}
This is the converse of deriving the BVP (\ref{eqpia}) and (\ref{eqpib}) from equation (\ref{defpi}) and the backward Kolmogorov equation.

\section{Multiple Brownian particles in an interval with global resetting}

Now suppose that there are $M$ non-interacting Brownian particles in the interval $[0,L]$ with totally absorbing boundaries at both ends. Let $X_m(t)$, $m=1,\ldots,M$, denote the position of the $m$th particle at time $t$ and let $\calT_m$ be the FPT for the $m$th particle to exit the domain through either end:
\begin{equation}
\calT_m=\inf\{t>0,\, X_m(t)\in \{0,L\}\}.
\end{equation}
For all $t<\calT$, where $\calT$ is the fastest FPT,
\begin{equation}
\calT=\min_m \{\calT_m\},
\end{equation}
we have the system of SDEs
\begin{eqnarray}
dX_m(t)&=\sqrt{2D}d{W_m}(t) + (x_{r,m}-X_m(t^-))dN(t),
\label{dXM}
\end{eqnarray}
where $W_m(t)$, $m=1,\ldots,M$ is a set of independent Wiener processes with
\begin{eqnarray}
\langle W_m(t)\rangle = 0,\quad \langle W_m(t)W_n(s)\rangle =\delta_{m,n}\min\{t,s\}.
\end{eqnarray}
We assume that the particles are subject to a global resetting protocol; at the sequence of Poisson-generated times $T_n$, $n\geq 1$, all the particles are simultaneously reset to their respective reset position $x_{r,m}$, $m=1,\ldots,M$. Let $\X(t)=(X_1(t),\ldots,X_M(t))$ and impose the initial condition $\X(0)=\x_0=(x_{0,1},\ldots,x_{0,m})$. Also set $\x_r=(x_{r,1},\ldots,x_{r,m})$. The multi-particle version of the generalised It\^o's lemma (\ref{lemma}) is
\begin{eqnarray}
\fl df(\X(t),t)&=&\bigg [\partial_t f(\X(t),t)+D\sum_{m=1}^M\partial^2_{x_m}f(\X(t),t)\bigg ]dt+\sqrt{2D}\sum_{m=1}^M\partial_{x_m} f(\X(t),t)dW_m(t)\nonumber \\
\fl &&\quad +\bigg [f(\x_r,t)-f(\X(t^-),t^-)\bigg ]dN(t).
\label{lemmaM}
\end{eqnarray}
Let $p^{(M)}(\x,t|\x_0)$ be the joint probability density for the $M$ particles to be at the positions $\X(t)=\x$ at time $t$ given the initial positions $\X(0)=\x_0$ and reset positions $\x_r$. Note that
\begin{eqnarray}
p^{(M)}(\x,t|\x_0)=\E[\delta(\x-\X(t)){\bf 1}_{\calT>t}|\X(0)=\x_0 ].
\end{eqnarray}
Introduce the survival probability that all $M$ particles are still in the interior $(0,L)$ up to time $t$:
\begin{equation}
\fl Q^{(M)}(\x_0,t)=\int_{\Omega}p^{(M)}(\x,t)d\x=\E[ {\bf 1}_{\calT>t}|\X(0)=\x_0 ],\quad \Omega =[0,L]^M.
\end{equation}
(Again we drop the explicit dependence of $p^{(M)}$ and $Q^{(M)}$ on $\x_r$.) 
The density $p^{(M)}$ satisfies the forward Kolmogorov equation 
\begin{eqnarray}
\fl \frac{\partial p^{(M)}(\x,t|\x_0)}{\partial t}=D\sum_{m=1}^M\frac{\partial^2 p^{(M)}(\x,t|\x_0)}{\partial x_m^2} +
r[\delta(\x-\x_r)Q^{(M)}(\x_0,t)-p^{(M)}(\x,t|\x_0)]  ,\nonumber \\ \fl
\label{FPresetM}
\end{eqnarray}
together with the initial condition $p^M(\x,0|\x_0)=\delta(\x-\x_0)$ and the Dirichlet boundary conditions
\begin{equation}
\left .p^{(M)}(\x,t|\x_0)\right |_{x_m=0,L}=0\mbox{ for all } m=1,\ldots,M.
\end{equation}
Here $\delta(\x-\x_0)=\prod_{m=1}^M\delta(x_m-x_{0,m})$ etc.
The set of boundary conditions means that $p^{(M)}$ vanishes as soon as one of the particles is absorbed. The structure of equation (\ref{FPresetM}) ensures that the joint probability density cannot be reduced to the product of single-particle densities \cite{Majumdar26}. That is,
\begin{equation}
\label{rnoteq}
 p^{(M)}(\x,t|\x_0)\neq \prod_{m=1}^Mp(x_m,t|x_{0,m} ),
\end{equation}
where $p$ evolves according to equation (\ref{FPreset}).  
 The relation (\ref{rnoteq}) reflects the fact that a global resetting protocol introduces statistical correlations between two or more particles, even in the absence of particle interactions. 
 
Let $\pi^{(M)}(\x_0;\x_r)$ denote the joint splitting probability that all $M$ particles are absorbed at the end $x=L$ given the initial positions $\X(0)=\x_0$ and reset positions $\x_r$. Proceeding previous studies of particles escaping from a stochastically gated domain \cite{Bressloff15a,Lawley16}, see also appendix A, we write down a BVP for $\pi^{(M)}$ and then use the generalised It\^o's lemma (\ref{lemmaM}) to show that $\pi^{(M)}$ has the correct physical interpretation. The BVP takes the form of a backward equation:
\begin{subequations}
 \begin{eqnarray}
0&=D\sum_{m=1}^M\frac{\partial^2 \pi^{(M)}(\x_0;\x_r)}{\partial x_{0,m}^2} +
r[\pi^{(M)}(\x_r)-\pi^{(M)}(\x_0;\x_r)] ,
\label{piresetM}
\end{eqnarray}
with $\pi^{(M)}(\x_r)=\pi^{(M)}(\x_0;\x_r)_{\x_0=\x_r}$, supplemented by the boundary conditions
\begin{equation}
\label{piBCa}
\left .\pi^{(M)}(\x_0;\x_r)\right |_{x_{0,m}=0}=0,\quad m=1,\ldots,M,
\end{equation}
and
\begin{eqnarray}
\label{piBCb}
\left . \pi^{(M)}(\x_0;\x_r)\right |_{x_{0,m}=L}=\pi^{(M-1)}(\x^{(m)}_0;\x^{(m)}_r) ,\quad m=1,\ldots,M.
\end{eqnarray}
\end{subequations}
We have set
\begin{equation}
\x_0^{(m)}=(x_{0,1},\ldots ,x_{0,m-1},x_{0,m+1},\ldots x_{0,m})
\end{equation}
and similarly for $\x_r^{(m)}$. Here $\pi^{(M-1)}$ denotes the corresponding splitting probability when the initial number of particles is $M-1$.

From the generalised It\^o's lemma (\ref{lemmaM}), we can write down the multi-particle version of equation (\ref{pilemma}) according to
\begin{eqnarray}
\fl d\pi^{(M)}(\X(t);\x_r)&=&\bigg [D\sum_{m=1}^M\partial^2_{x_m}\pi(\X(t);\x_r)\bigg ]dt+\sqrt{2D}\sum_{m=1}^M\partial_{x_m} \pi^{(M)}(\X(t);\x_r)dW_m(t)\nonumber \\
\fl &&\quad +\bigg [\pi^{(M)}(\x_r)-\pi^{(M)}(\X(t^-);\x_r)\bigg ]dN(t).
\label{pilemmaM}
\end{eqnarray}
Taking expectations of both sides yields
\begin{eqnarray}
\fl \frac{d\E[\pi^{(M)}(\X(t);\x_r)]}{dt}&=& \E\bigg [D\sum_{m=1}^M\partial^2_{x_m}\pi^{(M)}(\X(t);\x_r) +r [\pi^{(M)}(\x_r)-\pi^{(M)}(\X(t);\x_r)\bigg ] .\nonumber \\
\fl
\end{eqnarray}
Imposing equation (\ref{piresetM}) then shows that the right-hand side is zero, and we obtain the multi-particle version of equation (\ref{lemma2}):
\begin{equation}
\pi^{(M)}(\x_0;\x_r)=\E[\pi^{(M)}(\X(t);\x_r)].
\label{lemma2M}
\end{equation}
Suppose we have established that the solution $\pi^{M-1}$ is the splitting probability that all $M-1$ particles are absorbed at $x=L$. (This is certainly true when $M=2$, see section 2.) We will show that this result extends to the case of $M$ particles using proof by induction on the number of particles. First, set $t=\tau$ on the right-hand side of equation (\ref{lemma2M}), where
$\tau=\min_{m}\{\tau_m\}$ and $\tau_m$ is the FPT of the $m$th particle to be absorbed at either end. That is, $\tau$ is the time at which the first particle is absorbed. Again we are exploiting the strong Markov property. If the first particle is absorbed at $x=0$, then the boundary condition (\ref{piBCa}) implies that $\pi^{(M)}(\X(\tau);\x_r)=0$ and there is no contribution to the expectation on the right-hand side of (\ref{lemma2M}). On the other hand, if the first particle is absorbed at $x=L$ then we can use the boundary condition (\ref{piBCb}) and the induction hypothesis to infer that the right-hand side of equation (\ref{lemma2M}) is the splitting probability that all $M$ particles are absorbed at $x=L$. 

The structure of the BVP for $\pi^{(M)}$ given by equations (\ref{piresetM})--(\ref{piBCb}) 
means that 
\begin{equation}
\pi^{(M)}(\x_0;\x_r)\neq \prod_{m=1}^M\pi(x_{0,m};x_{r,m}),
\end{equation}
where $\pi(x_0;x_r)$ is the splitting probability that a single independent particle starting at $x_0$ and resetting to $x_r$ exits at $x=L$, see equations (\ref{spi1}) and (\ref{spi2}). Hence,  analogous to the NESS of a closed domain,, $\pi^{(M)}$ provides a time-independent probe for exploring the emergence of statistical correlations due to global resetting. (An alternative quantity is the time of the last particle to escape the domain, see the discussion.)

\section{Splitting probability for a pair of particles} In order to illustrate the emergent collective effects of global resetting, we will solve the BVP (\ref{piresetM})--(\ref{piBCb}) for a pair of Brownian particles with global resetting ($M=2$). For ease of notation, we set $\x_0=(x,y)$ and $\x_r=(x_r,y_r)$. We then have
\begin{subequations}
 \begin{eqnarray}
\fl 0&=D\frac{\partial^2 \pi^{(2)}(\x_0;\x_r)}{\partial x^2} +D\frac{\partial^2 \pi^{(2)}(\x_0;\x_r)}{\partial y^2}
+r[\pi^{(2)}(\x_r)-\pi^{(2)} (\x_0;\x_r] ,
\label{pireset2}
\end{eqnarray}
supplemented by the boundary conditions
\begin{equation}
\label{piBC2a}
 \pi^{(2)}(0,y;\x_r)=0=\pi^{(2)}(x,0;\x_r)
\end{equation}
and
\begin{eqnarray}
\label{piBC2b}
 \pi^{(2)}(L,y;\x_r)=\pi(y;y_r),\quad  \pi^{(2)}(x,L;\x_r)=\pi(x;x_r),\end{eqnarray}
\end{subequations}
with $\pi(x;x_r)$ the single-particle splitting probability given by equations (\ref{spi1}) and (\ref{spi2}).

We will solve the above inhomogeneous BVP in terms of the Dirichlet Green's function $G(x,y; X,Y)$ for the modified Helmholtz equation in a square. The latter satisfies 
\begin{subequations}
 \begin{eqnarray}
\frac{\partial^2 G}{\partial x^2} +\frac{\partial^2G}{\partial y^2}
-\xi^2 G=-\delta(x-X)\delta(y-Y),
\label{G}
\end{eqnarray}
with $\xi=\sqrt{r/D}$ as before and 
\begin{equation}
\fl G(0,y;X,Y)=G(x,0;X,Y)=G(L,y;X,Y)=G(x,L;X,Y)=0.
\label{BCG}
\end{equation}
\end{subequations}
The Green's function can be expanded in terms of the orthonormal eigenfunctions of the Laplacian in the square to give
\begin{eqnarray}
\fl G(x,y;X,Y) =\frac{4}{L^2}\sum_{n=1}^{\infty}\sum_{m=1}^{\infty}\frac{\sin(n \pi x/L)\sin(m \pi y/L)\sin(n \pi X/L)\sin(m \pi Y/L)}{\xi^2+(n\pi/L)^2+(m \pi/L)^2}.
\fl
\label{DG1}
\end{eqnarray}
This is a rapidly convergent series representation since the denominator grows quadratically with the integers $n,m$. A more computationally efficient representation can be obtained by Fourier transforming with respect to $x$, say, and solving the resulting one-dimensional problem in $y$. This yields the single summation
\begin{eqnarray}
\fl G(x,y;X,Y)=\frac{2}{L} \sum_{n=1}^{\infty} \frac{\sin(n \pi x/L) \sin(n \pi X/L)\sinh(\omega_n y_<)\sinh(\omega_n(L-y_{>}))}{\omega_n\sinh(\omega_nL)},
\label{DG2}
\end{eqnarray}
where
\begin{equation}
\omega_n=\sqrt{\xi^2+(n\pi /L)^2},
\end{equation}
and
\begin{eqnarray}
y_<=\min\{y,Y\},\quad y_>=\max\{y,Y\}.
\end{eqnarray}
Note that $G$ is symmetric, $G(x,y;X,Y)=G(X,Y;x,y)$.

We can now solve the inhomogeneous BVP for $\pi^{(2)}$ using Green's identity
\begin{eqnarray}
\fl \int_0^LdX \int_0^L dY G(X,Y;x,y)({\bm \nabla}^2-\xi^2)\pi^{(2)}(X,Y;\x_r)\nonumber \\
\fl -\int_0^LdX \int_0^L dY  \pi^{(2)}(X,Y;\x_r)({\bm \nabla}^2-\xi^2) G(X,Y;x,y) \\
\fl =\int_0^L dY \bigg [G(X,Y;x,y)\partial_X \pi^{(2)}(X,Y;\x_r)-\pi^{(2)}(X,Y;\x_r)\partial_X G(X,Y;x,y)\bigg ]_{X=0}^{X=L}\nonumber \\
\fl \quad + \int_0^L dX \bigg [G(X,Y;x,y)\partial_Y \pi^{(2)}(X,Y;\x_r)-\pi^{(2)}(X,Y;\x_r)\partial_Y G(X,Y;x,y)\bigg ]_{Y=0}^{Y=L}.\nonumber
\end{eqnarray}
Here
\begin{equation}
{\bm \nabla}^2=\frac{\partial^2}{\partial X^2}+\frac{\partial^2}{\partial Y^2}.
\end{equation}
Using equations (\ref{pireset2})--(\ref{BCG}), we find that
\begin{eqnarray}
\label{respi1}
\fl \pi^{(2)}(\x_0;\x_r)&=\xi^2 \pi^{(2)}(\x_r) \int_0^LdX \int_0^L dY G(X,Y;x,y)\\
\fl &\ -\int_0^LdY \pi(Y;y_r) \partial_X G(L,Y;x,y)-\int_0^LdX \pi(X;x_r) \partial_Y G(X,L;x,y).\nonumber 
\end{eqnarray}
Finally, setting $\x_0=\x_r$ we obtain a self-consistency condition for $\pi(\x_r)$ with the solution
\begin{equation}
\pi^{(2)}(\x_r)=-\frac{\int_0^LdY \pi(Y;y_r) \partial_X G(L,Y;x_r,y_r)+\int_0^LdX \pi(X;x_r) \partial_YG(X,L;x_r,y_r)}{1-\xi^2\int_0^LdX \int_0^L dY G(X,Y;x_r,y_r)}.
\label{respi2}
\end{equation}

Using the double summation (\ref{DG1}) and the symmetry property, we see that
\begin{eqnarray}
\fl \int_0^LdX \int_0^L dY G(X,Y;x,y) =\frac{16}{\pi^2}\sum_{{\rm odd }\ n} \sum_{{\rm odd} \ m}\frac{1}{nm}\frac{\sin(n \pi x/L)\sin(m \pi y/L)}{\xi^2+(n\pi/L)^2+(m \pi/L)^2}.\nonumber \\
\fl
\label{sum1}
\end{eqnarray}
In addition, integrating equation (\ref{G}) over the square $\Omega=[0,L]\times [0,L]$ and using the divergence theorem implies that
\begin{equation}
1-\xi^2 \int_{\Omega}   G(\x;{\bf X}) d\x =-\int_{\partial \Omega} {\bf n}(\x) \cdot {\bm \nabla}  G(\x;{\bf X}) d\x,
\end{equation}
where $\partial \Omega$ is the boundary of the square and ${\bf n}(\x)$ is the outward unit normal. The right-hand side of the above equation is the total exit flux from the square, which is positive definite. Hence, using the symmetry property of the Green's function, we deduce that the denominator of equation  (\ref{respi2}) is positive definite. Moreover, 
\begin{eqnarray}
\fl  \int_0^LdY \pi(Y;y_r) \partial_X G(L,Y;x,y)&=4 \sum_{n=1}^{\infty} \sum_{m=1}^{\infty} \frac{\sin(n \pi x/L)\sin(m \pi y/L)}{\xi^2+(n\pi/L)^2+(m \pi/L)^2}(-1)^nn\pi a_m(y_r),\nonumber \\
\fl &\equiv {\mathcal F}(x,y;y_r),
\label{sum2}
\end{eqnarray}
where 
\begin{eqnarray}
\fl a_m(y_r)&=\int_0^L\pi(y,y_r)\sin(m\pi y/L)dy\nonumber\\
\fl &=\int_0^L\bigg [\pi(x_r)(1-\cosh \xi y)]+\frac{1-\pi(x_r)(1-\cosh \xi L)}{\sinh\xi L}\sinh \xi y\bigg ]\sin(m\pi y/L)dy\nonumber \\
\fl &=\pi(x_r)\bigg [\frac{\cos m\pi}{m\pi/L}-\frac{m\pi/L}{\xi^2+(m\pi/L)^2}(1-\cos m\pi \cosh\xi L)\bigg ]\nonumber \\
\fl &\quad -\frac{1-\pi(x_r)(1-\cosh \xi L)}{\sinh\xi L}\frac{m\pi/L}{\xi^2+(m\pi/L)^2} \cos m \pi\sinh \xi L.
\end{eqnarray}
Similarly,
\begin{eqnarray}
\int_0^LdX \pi(X;x_r) \partial_Y G(X,L;x,y)={\mathcal F}(y,x;x_r).
 \end{eqnarray}
As expected, we find that $\partial_X G(L,Y;x,y)<0$ and $\partial_Y G(X,L;x,y)<0$ so $\pi^{(2)}(x,y)>0$ in equation (\ref{respi2}).

\begin{figure}[t!]
\centering
\includegraphics[width=10cm]{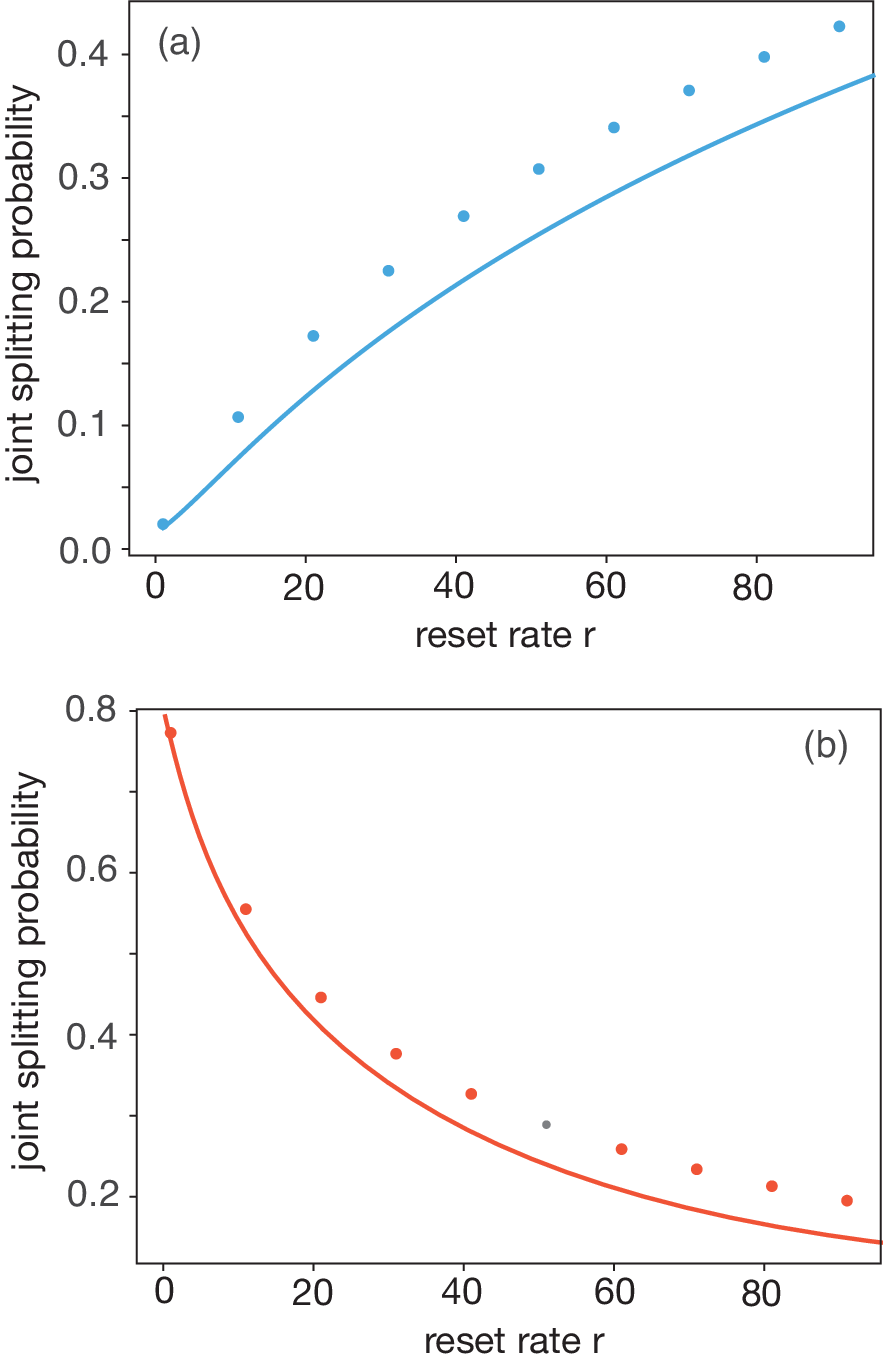}
\caption{Plots of $\pi^{(2)}(\x_0;\x_r)$ (global resetting) and $\pi^{(2)}_{\rm loc}(\x_0;\x_r)$ (local resetting) as functions of the resetting rate $r$, which are given by equations (\ref{respi1})--(\ref{respi2})) and (\ref{piloc}), respectively. We take $\x_0=(x,x)$ and $\x_r=(x_r,x_r)$ with (a) $x=0.1,x_r=0.9$ and (b) $x=0.9,x_r=0.1$.  The solid curves correspond to $ \pi_{\rm loc}^{(2)}$, whereas the discrete points sample $\pi^{(2)}$. We set $L=1$ and truncate the double summation of the Green's function in (\ref{DG1}) by taking $1\leq n,m\leq N$ with $N=1000$.}\label{fig2}
\end{figure}

\begin{figure}[t!]
\centering
\includegraphics[width=10cm]{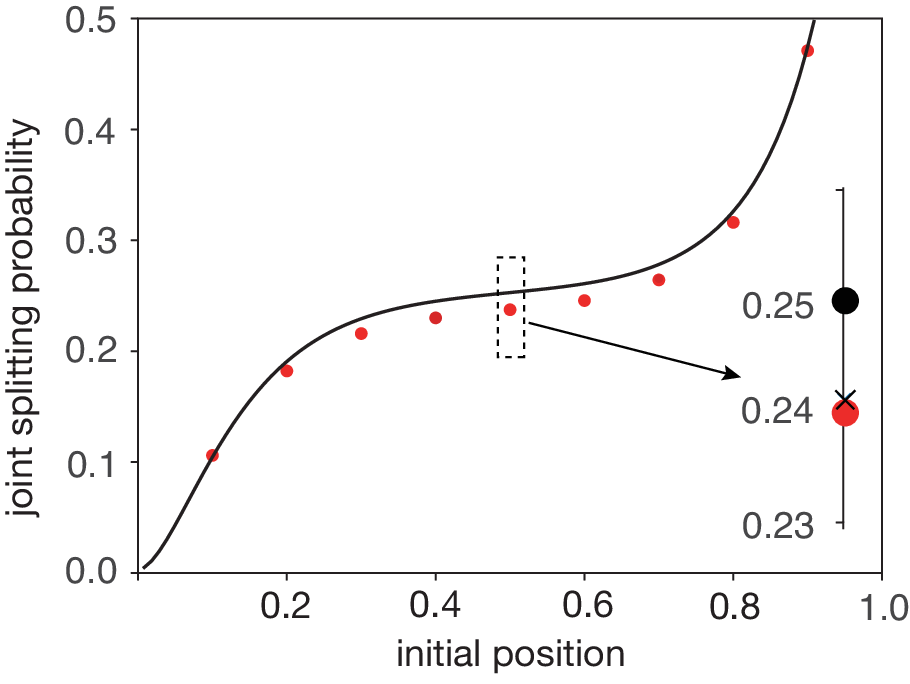}
\caption{Corresponding plots of $\pi^{(2)}(\x_0;\x_r)$ and $\pi^{(2)}_{\rm loc}(\x_0;\x_r)$ as functions of the initial position $x$ with $\x_0=(x,x)$ and $\x_r=(x_r,x_r)$. The solid curve corresponds to $ \pi_{\rm loc}^{(2)}$, whereas the discrete points sample $\pi^{(2)}$. We also set $L=1$, $r=100$ and $N=1000$.  The inset shows $\pi^{(2)}$ for $N=1000$ (solid red circle) and $N=1100$ (indicated by x) and $\pi^{(2)}_{\rm loc}$ (solid black circle) for $x=0.5$.}\label{fig3}
\end{figure}

We now compare the joint splitting probability $\pi^{(2)}(\x_0;\x_r)$ for a pair of particles with global resetting to the corresponding quantity with local resetting. In the latter case, each particle independently resets its position so that
\begin{equation}
 \pi_{\rm loc}^{(2)}(\x_0;\x_r) =\pi(x;x_r)\pi(y;y_r),
 \label{piloc}
\end{equation}
with $\pi(x;x_r)$ given by equations (\ref{spi1}) and (\ref{spi2}). For the sake of illustration, we assume the particles have the same initial position and the same reset position, that is, we set $y=x$ and $y_r=x_r$. We numerically evaluate the splitting probability with global resetting by truncating the double sums in equations (\ref{sum1}) and (\ref{sum2}) such that $1\leq n,m\leq N$ with $N=1000$. Although the effects of correlations are relatively small, they are statistically significant compared to truncation errors in the evaluation of the Green's functions. In Fig. \ref{fig2} we plot the splitting probabilities as a function of the resetting rate $r$ for the two cases $x=0.1,x_r=0.9$ and $x=0.9,x_r=0.1$. The solid curves correspond to $ \pi_{\rm loc}^{(2)}$, whereas the discrete points sample $\pi^{(2)}$. When $x<x_r$ resetting is clearly beneficial so the splitting probabilities are monotonically increasing functions of $r$. On the other hand, resetting is detrimental when $x>x_r$ so that the splitting probabilities are monotonically decreasing functions of $r$. In both cases we find that global resetting induces a small but statistically significant increase in the probability that both particles exit at the right-hand boundary. However, in other parameter regimes, global resetting can decrease the splitting probability as illustrated in Fig. \ref{fig3} for $x=x_r=0.5$.

\section{PDE perspective}
In this section we show how the BVP (\ref{piresetM})--(\ref{piBCb}) can be mapped into a hierarchy of moment equations for the following SPDE:
\begin{equation}
\label{SPDE}
\frac{\partial U(x,t)}{\partial t}=D\frac{\partial ^2U(x,t)}{\partial x^2}+[c(x)-U(x,t^-)]dN(t),
\end{equation}
supplemented by the inhomogeneous boundary conditions
\begin{equation}
U(0,t)=0,\quad U(L,t)=\eta >0.
\end{equation}
Here $U(x,t)$ is a stochastic concentration field that is instantaneously reset to a prescribed concentration $c(x)$ according to the Poisson process $N(t)$. Note that particle number is not conserved due to the presence of the source term at $x=L$ and due to stochastic resetting. In the latter case, the number of particles resets to a fixed amount ${\mathcal N}_0=\int_0^Lc(x)dx$.

We will derive moment equations for $U(x,t)$ using the method introduced in Ref. \cite{Bressloff15a} for the stochastically gated diffusion equation. The first step is to spatially discretise the SPDE (\ref{SPDE}) using a finite-difference scheme. One of the nice features of this discretisation is that we can incorporate the boundary conditions into the resulting discrete Laplacian. Introduce the lattice spacing $a$ such that $(N+1)a=L$ for integer $N$, and
let $U_k(t)=U(ak,t)$, $k=0,\ldots, N+1$. Similarly, set $c_k=c(ak)$. Then
\begin{equation}
\label{sh}
\fl \frac{dU_k(t)}{dt}=\sum_{l=1}^N\Delta_{kl}U_l(t) +\eta_a \delta_{k,N}+[c_{k}-U_k(t^-)]\frac{dN(t)}{dt},\quad k=1,\ldots, N,\quad 
\end{equation}
with $\eta_a={\eta D}/{a^2} $.
Away from the boundaries ($k\neq 1,N$), $\Delta_{kl}$ is given by the discrete Laplacian
\begin{subequations}
\begin{equation}
\label{dL1}
\Delta_{kl}=\frac{D}{a^2}[\delta_{k,l+1}+\delta_{k,l-1}-2\delta_{k,l}],
\end{equation}
whereas
\begin{equation}
\label{dL2}
 \Delta_{1l}=\frac{D}{a^2}[\delta_{l,2}-2\delta_{l,1}], \quad \Delta_{Nl}=\frac{D}{a^2}[\delta_{N-1,l}-2\delta_{N,l}].
\end{equation}
\end{subequations}
Equation (\ref{sh}) is a finite-dimensional differential equation with global stochastic resetting.
  Let $\U(t)=(U_1(t),\ldots,U_N(t))$ and introduce the probability density 
\begin{equation}
 \mbox{Prob}\{\U(t)\in (\u,\u+d\u)\}=p(\u,t)d\u .
\end{equation}
The corresponding forward Kolmogorov equation is
\begin{equation}
\label{CK0}
\fl \frac{\partial p(\u,t)}{\partial t}=-\sum_{k=1}^N\frac{\partial}{\partial u_k}\left [\left (\sum_{l=1}^N\Delta_{kl}u_l+\eta_a \delta_{k,N} \right )p(\u,t)\right ]+r[p_0(\u)-p(\u,t)],
\end{equation}
where 
\begin{equation}
p_0(\u)=\prod_{k=1}^N\delta(c_{k}-u_k).
\end{equation}
The first-order derivatives represent Liouville terms associated with the deterministic flow, whereas the final term on the right-hand side is due to resetting. (If we added spatio-temporal white noise terms on the right-hand side of the SPDE (\ref{SPDE}) then there would be additional second-order derivative terms in equation (\ref{CK0}).)  

Introduce the first-order moments
\begin{equation}
\calU_{k}(t)=\E[U_k(t)]=\int p(\u,t)u_kd\u.
\end{equation}
Multiplying both sides of the Kolmogorov equation (\ref{CK0}) by $u_k(t)$ and integrating with respect to $\u$ gives (after integrating by parts and using that $p(\u,t)\rightarrow 0$ as $\u\rightarrow \infty$ by the maximum principle)
\begin{equation}
\frac{d \calU_{k}(t)}{d t}=\sum_{l=1}^N\Delta^n_{kl}\calU_{l}(t)+\eta_a\delta_{k,N}+ r[c_{k}-\calU_k(t)].
\end{equation}
If we now retake the continuum limit $a\rightarrow 0$, $N\rightarrow \infty$ and define
\begin{equation}
\label{vdirichlet}
\calU(x,t)=\E[U(x,t)],
\end{equation}
we obtain the continuum first-order moment equation
\begin{eqnarray}
\label{CKdirichlet}
\frac{\partial \calU(x,t)}{\partial t} &=D\frac{\partial^2 \calU(x,t)}{\partial x^2}+r[c(x)-\calU(x,t)],
\end{eqnarray}
with $\calU(0,t)=0$ and $\calU(L,t)=  \eta$.
If $\eta >0$ then there exists a nontrivial stationary solution satisfying
\begin{eqnarray}
\label{NESS1}
0&=D\frac{d^2 \calU(x)}{d x^2}+r[c(x)-\calU(x)],
\end{eqnarray}
with $\calU(0)=0$ and $ \calU(L)=  \eta$. Integrating boths sides of (\ref{NESS1}) with respect to $x$ implies that
\begin{equation}
\frac{1}{r}\bigg [D\partial_x\calU(0)-D\partial_x\calU(L)\bigg ]={\mathcal N}_0-\int_0^L\calU(x)dx.
\end{equation}
Each time the system resets, there is a jump in the number of particles given by the right-hand side This must be balanced by the mean exit flux times the mean time between resets, which is given by the left-hand side.
If we take $c(x)=c_r$, a constant, then equation (\ref{NESS1}) is formally identical to the single-particle BVP for the splitting probability $\pi(x_0)$, see equations (\ref{eqpia}) and (\ref{eqpib}), under the mapping:
\begin{equation}
\calU(x)\rightarrow \eta \pi(x) .
\end{equation}
One major difference from the single-particle model is that $c_r$ is a prescribed constant rather than an unknown solution $\pi(x_r)$ that has to be determined self-consistently. We now show that an analogous result holds for the higher-order moments.

First, consider the second-order moments of the spatially discretised system
\begin{equation}
\calU^{(2)}_{kl}(t)=\E[U_k(t)U_l(t)]=\int p(\u,t)u_k(t)u_l(t)d\u.
\end{equation}
Multiplying both sides of the Kolmogorov equation (\ref{CK0}) by $u_k(t)u_l(t)$ and integrating with respect to $\u$ gives (after integration by parts)
\begin{align}
\frac{d \calU^{(2)}_{kl}}{d t}=\sum_{j=1}^N\left [\Delta_{kj}\calU^{(2)}_{jl}+\Delta_{lj}\calU^{(2)}_{jk}\right ]+\eta_a \left [\calU_{k}\delta_{l,N}+ \calU_{l}\delta_{k,N}\right ]+r[c_{k}c_{l}-\calU^{(2)}_{kl}].\end{align}
In the continuum limit $a\rightarrow 0$ and $N\rightarrow \infty$, we obtain the continuum second-order moment equation for \begin{equation}
\label{v2dirichlet}
\calU^{(2)}(x,y,t)=\E[U(x,t)U(y,t)],
\end{equation}
which is given by
  \begin{eqnarray}
    \label{CK2dirichlet}
  \frac{\partial \calU^{(2)}}{\partial t} &=D\frac{\partial^2 \calU^{(2)}}{\partial x^2}+D\frac{\partial^2 \calU^{(2)}}{\partial y^2}+r[c(x)c(y)-\calU^{(2)}],
  \end{eqnarray}
together with the boundary conditions
\begin{subequations}
\begin{equation}
\calU^{(2)}(0,y,t)=\calU^{(2)}(x,0,t)=0,
\end{equation}
and
\begin{equation}
\calU^{(2)}(L,y,t)=\eta \calU(y,t),\quad \calU^{(2)}(x,L,t)=\eta \calU(x,t).
\end{equation}
\end{subequations}
Physically speaking, $\calU^{(2)}(x,y,t)$ represents the two-point equal time correlation function of the stochastic field $U(x,t)$.
 We can proceed along similar lines to derive the Kolmogorov equation
for the $M$th order moment function in the continuum limit, which is defined according to
\begin{equation}
\label{vn}
\calU^{(M)}(\x,t)=\E[U(x_1,t)U(x_2,t)\ldots U(x_M,t)],
\end{equation}
with $ \x=(x_1,\ldots ,x_m)$. We find that
\begin{subequations}
  \begin{eqnarray}
    \label{CKn}
  \frac{\partial \calU^{(M)}(\x,t)}{\partial t} &=D\sum_{m=1}^M\frac{\partial^2 \calU^{(M)}(\x,t)}{\partial x_m^2}
  +r[c^{(M)}(\x)-\calU^{(M)}(\x,t)],
  \end{eqnarray}
where $c^{(M)}(\x)=\prod_{m=1}^Mc(x_m),$
supplemented by the boundary conditions
\begin{equation}
\label{inhoma}
\left . \calU^{(M)}(\x,t) \right |_{x_m=0} =0,
\end{equation}
and
\begin{equation}\label{inhomb}
\left . \calU^{(M)}(\x,t) \right |_{x_m=L}= \eta \calU^{(M-1)}(x_1,\ldots,x_{m-1},x_{m+1}\ldots,x_M,t)
\end{equation}
\end{subequations}
for $m=1,\ldots,M$. 

As in the case of the first-order moment equation, the boundary source terms ensure that there exists a unique stationary solution of the $M$th-order moment equation, which depends on the stationary solution of the $(M-1)$-th moment equation via the boundary condition (\ref{inhomb}). That is, we can take the large-$t$ limit
\begin{equation}
 \calU^{(M)}(\x)=\lim_{t\rightarrow \infty}  \calU^{(M)}(\x,t).
 \end{equation}
Moreover, if $c(x)=c_r$ then $c^{(M)}(\x)=c_r^M$ and the time-independent version of equations (\ref{CKn})--(\ref{inhomb}) is formally identical to the BVP (\ref{piresetM})--(\ref{piBCb}) for the splitting probability $\pi^{(M)}$ under the mapping 
\begin{equation}
\calU^{(M)}(\x)\rightarrow   \eta^M \pi^{(M)}(\x).
\end{equation}
However, the constant $c_r^M$ is a prescribed constant constrained by the product structure of the reset concentration rather than the $M$th-order version of the unknown solution $\pi^{(M)}(\x_r)$.

\section{Discussion} In this paper we extended a previous study of the exit statistics for two or more Brownian particles in a stochastically gated domain \cite{Bressloff15a,Lawley16} to the case of Brownian particles subject to global resetting. We focused on the splitting probability $\pi^{(M)}$ that $M$ initial particles all exit from the same boundary in an open interval. In particular, we showed the following: (i) the inhomogeneous BVP for $\pi^{(M)}$ couples to the corresponding BVP for $\pi^{(M-1)}$, thus forming a nested sequence. (ii) The BVP for $\pi^{(M)}$ can be mapped to the $M$th order stationary moment equation for the diffusion equation driven by stochastic resetting. We also illustrated the theory by solving the BVP for a pair of particles ($M=2$) and established that correlations due to global resetting yield a small but statistically significant effect on the joint splitting probability.

A few general comments are in order. First, the probabilistic methods used to determine the BVP for $\pi^{(M)}$ can also be used to obtain the BVP of other quantities characterising the effects of globally resetting on the exit statistics of multiple particles. One example is the mean first passage time (MFPT) of the last particle to exit the interval.
Suppose that the function $\calT^{(M)}(\x_0;\x_r)$ satisfies the BVP
\begin{subequations}
 \begin{eqnarray}
-1&=D\sum_{m=1}^M\frac{\partial^2 \calT^{(M)}(\x_0;\x_r)}{\partial x_{0,m}^2} +
r[\calT^{(M)}(\x_r)-\calT^{(M)}(\x_0;\x_r)] 
\label{TM}
\end{eqnarray}
with $\calT^{(M)}(\x_r)=\calT^{(M)}(\x_0;\x_r)_{\x_0=\x_r}$, supplemented by the boundary conditions
\begin{eqnarray}
\label{TBCb}
\left . \calT^{(M)}(\x_0;\x_r)\right |_{x_{0,m}=0,L}=\calT^{(M-1)}(\x^{(m)}_0;\x^{(m)}_r) ,\quad m=1,\ldots,M,
\end{eqnarray}
\end{subequations}
with $\calT^{(0)}\equiv 0$. Let ${\mathcal S}^{(M)}=\sup_{1\leq m \leq M}\calT_m$ be the FPT of the last particle given $\X(0)=\x_0$. Following an analogous proof to Ref. \cite{Lawley16}, it can be shown that the solution of the above BVP satisfies the relation
\begin{equation}
\calT^{(M)}(\x_0;\x_r)=\E[{\mathcal S}^{(M)}].
\end{equation}
Second, although we restricted our analysis to exit from an interval, the same methods could be applied to exit from a higher-dimensional domain. Finally, it should be possible to apply the methods of this paper to any system of non-interacting particles subject to a globally fluctuating environment, provided that there exists an underlying generalised It\^o's lemma. Hence, we could consider active run-and-tumble particles rather than Brownian particles \cite{Bressloff25}, and globally fluctuating potentials rather than stochastically gated boundaries or global resetting. However, a major challenge in all of these cases is solving the resulting sequence of nested BVPs, which become increasingly difficult as $M$ increases. It would also be of interest to explore whether some form of extreme statistics holds when $M\rightarrow \infty$.

\setcounter{equation}{0}
\renewcommand{\theequation}{A.\arabic{equation}}
\section*{Appendix A: Brownian particles in a stochastically gated interval}

In this appendix we summarise various results that were previously obtained for multiple particles exiting from a stochastically gated interval \cite{Bressloff15a,Lawley16}. We thus highlight the general nature of the mathematical framework used in this paper to analyse dynamically emergent correlations in exit statistics. 

\subsection*{A.1 Single particle} First, consider a single Brownian particle in the interval $[0,L]$ with an open boundary at $x=0$ and a stochastically gated boundary at $x=L$ whose state evolves according to the two-state Markov chain (\ref{M}). Let $p_{ij}(x,t|x_0)$ denote the joint probability density that $X(t)=x$ and $J(t)=i$ given the initial conditions $X(0)=x_0$ and $J(0)=j$. The forward Kolmogorov equation takes the form
 \begin{subequations} \begin{eqnarray}
  \label{sforwarda}
  \frac{\partial p_0}{\partial t} &=D\frac{\partial^2 p_0}{\partial x^2}-\beta p_0+\alpha p_1 ,\\
   \frac{\partial p_1}{\partial t} &= D \frac{\partial^2p_1}{\partial x^2 }+\beta p_0-\alpha p_1,
   \label{sforwardb}
  \end{eqnarray}
  \end{subequations}
with the boundary conditions
\begin{equation}
p_0(0,t)=p_1(0,t)=0,\quad p_0(L,t)=0,\quad \partial_xp_1(L,t)=0.
\end{equation}
We have used the shorthand notation $p_i(x,t)=p_{ij}(x,t|x_0,0)$ for $x_0,j$ fixed. The splitting probability that the particle is absorbed at the end $x=L$ is given by
\begin{equation}
\label{gam}
\pi_j(x_0)=-D\int_0^{\infty}\frac{\partial p_{0j}(L,\tau|x_0)}{\partial x}d\tau,
\end{equation}
which is the total probability flux through $x=L$ integrated over the time interval $[0,\infty)$.
One way to determine $\pi_j(x_0)$ is to consider 
the corresponding backward Kolmogorov equation for $q_i(x_0,t)= p_{0j}(x,t|x_0,0)$ with $x$ fixed:
\begin{subequations}
\begin{eqnarray}
  \label{sbackward1}
  \frac{\partial q_0}{\partial t} &=&D\frac{\partial^2 q_0}{\partial x_0^2} - \beta [q_0-  q_1], \\
  \label{sbackward2}
   \frac{\partial q_1}{\partial t} &=&D\frac{\partial^2 q_1}{\partial x_0^2} + \alpha [q_0-  q_1].
   \end{eqnarray}
    \end{subequations}
Differentiating equations (\ref{sbackward1}) and (\ref{sbackward2}) with respect to $x$ and integrating with respect to $t$, we find that
\begin{eqnarray}
 \label{split1}
0 &=&D\frac{d^2 \pi_0}{d x_0^2} - \beta [\pi_0-  \pi_1],\\
0 &=&D\frac{d^2 \pi_1}{d x_0^2} + \alpha [\pi_0-  \pi_1],
\label{split2}
\end{eqnarray}
with boundary conditions
\begin{equation}
\pi_{0}(0)=\pi_{1}(0)=0,\quad \pi_{0}(L)=1,\quad \partial_{x_0} \pi_1(L)=0.
\end{equation}

The physical quantity of interest is the total splitting probability for exit at $x=L$ irrespective of the initial state of the gate, which is given by
\begin{equation}
\pi(x_0)=\rho_0 \pi_0(x_0)+\rho_1\pi_1(x_0),
\end{equation}
with
\begin{equation}
\rho_0=\frac{\alpha}{\alpha+\beta},\quad \rho_1 =\frac{\beta}{\alpha+\beta}.
\end{equation}
Introduceing the scalings $\widehat{\pi}_0(x_0)=\rho_0\pi_0(x_0)$ and $\widehat{\pi}_1(x_0)=\rho_1 \pi_1(x_0)$, we have
\begin{subequations}
\begin{eqnarray}
 \label{Split1}
0 &=&D\frac{d^2 \widehat{\pi}_0}{d x_0^2} - \beta \widehat{\pi}_0+\alpha \widehat{\pi}_1,\\
0 &=&D\frac{d^2 \widehat{\pi}_1}{d x_0^2} + \beta \widehat{\pi}_0-  \alpha \widehat{\pi}_1
\label{Split2}
\end{eqnarray}
and
\begin{equation}
\widehat{\pi}_{0}(0)=\widehat{\pi}_{1}(0)=0,\quad \widehat{\pi}_{0}(L)=\rho_0,\quad \partial_y \widehat{\pi}_1(L)=0.
\end{equation}
 \end{subequations}
 Adding equations (\ref{Split1}) and (\ref{Split2}) and using the boundary conditions gives
\begin{equation}
\frac{d^2 {\pi} }{d x_0^2}=0,\quad 
\pi(0)=0,\quad {\pi}(L)=\rho_0+\widehat{\pi}_1(L),
\end{equation}
where $\widehat{\pi}_1(L)$ is an undetermined constant.
It follows that
\[{\pi}(x_0)=\frac{x_0}{L}[\rho_0+\widehat{\pi}_1(L)],\]
with
\begin{equation}
D\frac{d^2\widehat{\pi}_1}{dx_0^2}-(\alpha+\beta)\widehat{\pi}_1=-\frac{\beta}{L} x_0(\rho_0+\widehat{\pi}_1(L))
\end{equation}
and $\widehat{\pi}_1(0)=0,\partial_x\widehat{\pi}_1(L)=0$. We thus obtain the implicit solution
\begin{equation}
\label{V1}
\widehat{\pi}_1(x_0)=\rho_1(\rho_0+\widehat{\pi}_1(L))\left [-\frac{1}{\xi L}\frac{\sinh(\xi x_0)}{\cosh(\xi L)}+\frac{x_0}{L}\right ],
\end{equation}
where $\xi=\sqrt{(\alpha+\beta)/D}$.
Finally, setting $x_0=L$ we obtain a self-consistency condition for the unknown $\widehat{\pi}_1(L)$, which can be rearranged to give
\[\widehat{\pi}_1(L)=\rho_1\frac{1-(\xi L)^{-1}\tanh(\xi L)}{1+(\rho_1/\rho_0)(\xi L)^{-1}\tanh(\xi L)}\]
We thus find that the unconditional splitting probability is \cite{Bressloff15a}
\begin{equation}
\pi(x_0)=\frac{x_0}{L}\frac{1}{{1+(\rho_1/\rho_0)(\xi L)^{-1}\tanh(\xi L)}},
\label{gpi}
\end{equation}
whereas
 \begin{subequations}
 \begin{equation}
\label{gpi0}
\widehat{\pi}_0(x_0)=\rho_0\Gamma(\xi L) \left [\frac{\beta}{\alpha}\frac{1}{\xi L}\frac{\sinh(\xi x_0)}{\cosh(\xi L)}+\frac{x_0}{L}\right ],
\end{equation}
and
\begin{equation}
\label{gpi1}
\widehat{\pi}_1(x_0)=\rho_1 \Gamma(\xi L)\left [-\frac{1}{\xi L}\frac{\sinh(\xi x_0)}{\cosh(\xi L)}+\frac{x_0}{L}\right ].
\end{equation}
 \end{subequations}
 We have also set
 \begin{equation}
 \Gamma(s)=\frac{1}{{1+(\rho_1/\rho_0) \tanh(s)/s} }.
 \end{equation}

\subsection*{A.2 Multiple particles} 

Suppose that there are now $M$ non-interacting Brownian particles in the stochastically gated interval. Let $\X(t)=(x_1(t),\ldots x_M(t))$ denote the positions of the $M$ Brownian particles at time $t$ given the initial positions $\X(0)=\x_0$. Similarly, let $J(t)$ denote that state of the state given $J(0)=j$. We are interested in the splitting probability $\pi^{(M)}_j(x_{0,1},\ldots x_{0,M})$ that all $M$ particles are absorbed at the gated end $x=L$. We will give a simplified version of the proof presented in Ref. \cite{Lawley16} that $\pi_j^{(M)}$ is the solution of the following boundary value problem (BVP):
 \begin{subequations}
   \begin{eqnarray}
  \label{gbackwarda}
0 &=D\sum_{m=1}^M\frac{\partial^2 \pi^{(M)}_0}{\partial x_{0,m}^2}-\beta [\pi^{(M)}_0-\pi^{(M)}_1 ]\\
 0 &= D\sum_{m=1}^M\frac{\partial^2\pi^{(M)}_1}{\partial x_{0,m}^2 }+\alpha[ \pi^{(M)}_0-\pi^{(M)}_1],
   \label{gbackwardb}
  \end{eqnarray}
supplemented by the boundary conditions
\begin{equation}
\label{gBCa}
\left .\pi_j^{(M)}(\x_0)\right |_{x_{0,m}=0}=0, \quad j =0,1,
\end{equation}
and
\begin{eqnarray}
\label{gBCb}
\fl \left . \pi_0^{(M)}(\x_0)\right |_{x_{0,m}=L}=\pi_0^{(M-1)}(\x_0^{(m)}) ,\quad \left . \partial_{x_{0,m}}\pi_1^{(M)}(\x_0)\right |_{x_{0,m}=L}=0
\end{eqnarray}
 \end{subequations}
for $m=1,\ldots,M$.
Here $D$ is the diffusivity of each particle, and 
\begin{equation}
\fl \x_0=(x_{0,1},\ldots x_{0,M}),\quad \x_0^{(m)}=(x_{0,1},\ldots ,x_{0,m-1},x_{0,m+1},\ldots x_{0,M}).
\end{equation}
We proceed along analogous lines to the proof developed in section 3 for global resetting. 

Introducing an arbitrary test function $f(\X(t),J(t),t)$ we have the generalised It\^o's lemma
\begin{eqnarray}
\label{lemmaMg}
\fl df(\X(t),J(t),t)&=\bigg [\partial_t f(\X(t),J(t),t)+D\sum_{m=1}^M\partial^2_{x_m}f(\X(t),J(t),t)\bigg ]dt\\
\fl &\quad +\sqrt{2D}\sum_{m=1}^M\partial_{x_m} f(\X(t),J(t),t)dW_m(t)+\sum_{i=0,1}Q_{J(t)i} f(\X(t),i,t)dt,\nonumber 
\end{eqnarray}
where ${\bf Q}$ is the matrix generator of the two-state Markov chain:
\begin{equation}
{\bf Q}=\left (\begin{array}{cc} -\beta &\beta  \\ \alpha & -\alpha \end{array}\right ).
\end{equation}
Rewriting $\pi^{(M)}_{j}(\x_0)$ as $\pi^{(M)}(\x_0,j)$, applying It\^o's lemma to $\pi^{(M)}(\X(t),J(t))$ and then taking expectations with respect to the Gaussian white noise processes and the stochastic gate gives
\begin{eqnarray}
\fl \frac{d\E[\pi^{(M)}(\X(t),J(t))]}{dt}&=& \E\bigg [D\sum_{m=1}^M\partial^2_{x_m}\pi^{(M)}(\X(t),J(t)) +\sum_{i=0,1} Q_{J(t)i}\pi^{(M)}(\X(t),i) \bigg ] .\nonumber \\
\fl
\end{eqnarray}
Using the explicit form for ${\bf Q}$, and imposing equations (\ref{gbackwarda})--(\ref{gbackwardb}) shows that the right-hand side is zero. We thus obtain the analog of equation (\ref{lemma2M}), namely,
\begin{equation}
\pi^{(M)}(\x_0,j)=\E[\pi^{(M)}(\X(t), J(t))].
\label{lemma2Mg}
\end{equation}
Let $\calT$ be the FPT for the first particle to exit the interval at either end and set $t=\calT$ in equation (\ref{lemma2Mg}). (We again make use of the strong Markov property.) Clearly if the particle exits at $x=0$ then the boundary condition (\ref{gBCa}) implies that $\pi^{(M)}(\X(\tau), J(\tau))=0$ irrespective of the state of the gate $J(\tau)$. On the other hand, if the particle exits at $x=L$ then by construction $J(\tau)=0$ and the first boundary condition in (\ref{gBCb}) applies. Combining this with the induction hypothesis proves that the right-hand side of equation (\ref{lemma2Mg}) is the splitting probability that all $M$ particles are absorbed at $x=L$. 

The calculation of $\pi_j^{(M)}$ is non-trivial even for $M=2$. Here we simply sketch the main steps; the full details can be found in Ref. \cite{Bressloff15a}. For ease of notation we write $\x_0=(x,y)$. As in the single-particle case, the unconditional splitting probability is
\begin{equation}
\pi^{(2)}(x,y)=\rho_0 \pi_0^{(2)}(x,y)+\rho_1 \pi_1^{(2)}(x,y).
\end{equation}
Setting $\widehat{\pi}^{(2)}_j(\x_0)=\rho_j\pi^{(2)}_j(\x_0)$, $j=0,1$, we have
\begin{subequations}
  \begin{eqnarray}
    \label{CK2}
 0&=D\frac{\partial^2 \widehat{\pi}^{(2)}_0}{\partial x^2}+D \frac{\partial^2 \widehat{\pi}^{(2)}_0}{\partial y^2}-\beta \widehat{\pi}^{(2)}_0+\alpha \widehat{\pi}^{(2)}_1, \\
    \label{CK21}
  0&= D\frac{\partial^2\widehat{\pi}^{(2)}_1}{\partial x^2 }+D\frac{\partial^2\widehat{\pi}^{(2)}_1}{\partial y^2 }+\beta \widehat{\pi}^{(2)}_0-\alpha \widehat{\pi}^{(2)}_1,
  \end{eqnarray}
supplemented by the boundary conditions
\begin{equation}
\label{BC1}
\widehat{\pi}^{(2)}_j(0,y)=0=\widehat{\pi}^{(2)}_j(x,0),\quad j=0,1,
\end{equation}
\begin{equation}
\label{BC2}
\fl \widehat{\pi}^{(2)}_0(L,y)=  \widehat{\pi}_0(y),\, \widehat{\pi}^{(2)}_0(x,L)=   \widehat{\pi}_0(x), 
\end{equation}
and
\begin{equation}
\partial_x\widehat{\pi}^{(2)}_1(L,y)=\partial_y\widehat{\pi}^{(2)}_1(x,L)=0.
\label{BC3}
\end{equation}
\end{subequations}
Here $\widehat{\pi}_0(x)$ and $\widehat{\pi}_1(x)$ are given by equations (\ref{gpi0}) and (\ref{gpi1}), respectively.
Adding the pair of equations (\ref{CK2}) and (\ref{CK21}) yields
\begin{subequations}
\begin{equation}
\frac{\partial^2 {\pi}^{(2)}}{\partial x^2}+\frac{\partial^2 \pi^{(2)}}{\partial y^2}=0,
\end{equation}
with boundary conditions
\begin{equation}
{\pi}^{(2)}(0,y)={\pi}^{(2)}(x,0)=0,
\end{equation}
and
\begin{equation}
\label{CVC}
{\pi}^{(2)}(L,y)= \widehat{\pi}_0(y)+\widehat{\pi}^{(2)}_1(L,y),\quad {\pi}^{(2)}(x,L)=\widehat{\pi}_0(x)+\widehat{\pi}^{(2)}_1(x,L).
\end{equation}
\end{subequations}
Using separation of variables and the boundary conditions (\ref{BC1}) leads to the general solution 
\begin{eqnarray}
\label{genC}
\fl {\pi}^{(2)}(x,y)=\frac{A_0}{L^2}xy+\sum_{n>0}A_n[\sinh(n\pi x/L)\sin(n \pi y/L)+\sin(n\pi x/L)\sinh(n\pi y/L)].\nonumber \\
\end{eqnarray}
Moreover, the equation for $\widehat{\pi}^{(2)}_1$ can be rewritten as
  \begin{eqnarray}
 \frac{\partial^2\widehat{\pi}^{(2)}_1}{\partial x^2 }+\frac{\partial^2\widehat{\pi}^{(2)}_1}{\partial y^2 }-(\alpha+\beta)\widehat{\pi}^{(2)}_1(x,y)=-\beta {\pi}^{(2)}(x,y).
  \end{eqnarray}
Imposing the boundary conditions (\ref{BC1}) for $j=1$ leads to the general solution  
\begin{eqnarray}
\label{C1}
\widehat{\pi}^{(2)}_1(x,y)&=\rho_1  {\pi}^{(2)}(x,y)+B_0 \left [\frac{y}{L}\sinh (\xi x)+\frac{x}{L}\sinh(\xi  y)\right ]\\
&\quad +\sum_{n>0}B_n\sinh(\sqrt{(n\pi /L)^2+\xi^2}x)\sin(n \pi y/L)\nonumber \\
&\quad +\sum_{n>0}B_n\sin(n\pi x/L)\sinh(\sqrt{(n\pi /L)^2+\xi^2}y).\nonumber
\end{eqnarray}
Finally, the unknown coefficients are obtained by imposing the various boundary conditions at $x=L$. (By symmetry the boundary conditions at $y=L$ are automatically satisfied.) The first equation in (\ref{CVC}) yields, after equating coefficients in the Fourier series expansion with respect to $y$,
\begin{subequations}
\begin{equation}
A_0= \Gamma(\xi L)\bigg [ 1-\frac{\rho_1}{\rho_0}\frac{ \tanh \xi L}{\xi L}\bigg ],\quad B_0=-\frac{\rho_1 \Gamma(\xi L)}{\xi L\cosh(\xi L)}\end{equation}
and
\begin{equation} \rho_0A_n\sinh(n\pi)=B_n\sinh(\sqrt{(n\pi /L)^2+\xi^2}L),\quad n >0.\end{equation}
\end{subequations}

The final step is to determine the coefficients $A_n,n>0$ using the boundary condition $\partial_x\pi^{(2)}_1(L,y)=0$. After some long calculations we obtain the following infinite-dimensional matrix equation \cite{Bressloff15a}:
\begin{eqnarray}\label{hat}
\fl &\widehat{\gamma}_m{A}_m\sinh(m\pi)\nonumber \\
\fl & \quad +\sum_{n>0}(-1)^{n+m+1}\Big [\frac{\rho_1n}{n^2+m^2}+ \frac{\rho_0 n}{n^2+(\xi L/\pi)^2+m^2}\Big ] \sinh(n\pi){A}_n=-\frac{\Lambda_m}{m},\end{eqnarray}
where
\begin{eqnarray*}
\fl \widehat{\gamma}_m&=\frac12\left [\rho_1\pi \mbox{coth}(m\pi)+\rho_0\sqrt{\pi^2+(\xi L/m)^2}\mbox{cotanh}(\sqrt{(m\pi)^2+(\xi L)^2})\right ],
\end{eqnarray*}
and
\begin{eqnarray*}
\fl  \Lambda_m&=(\xi L\cosh(\xi L)B_0+\rho_1 A_0 )\frac{1}{m\pi}(-1)^{m+1}+(-1)^{m+1}B_0\frac{m\pi }{\xi^2+(m\pi/L)^2}\sinh(\xi L).
 \end{eqnarray*}
It can be checked that $A_m\sim \e^{-m\pi}/m^2$ for large $m$, and that the numerical solution of a truncated version of the matrix equation converges to a unique solution, except for a small boundary layer around $x=L$, which shrinks as more terms in the numerical approximation scheme are included \cite{Bressloff15a}. 

\begin{figure}[t!]
\centering
\includegraphics[width=12cm]{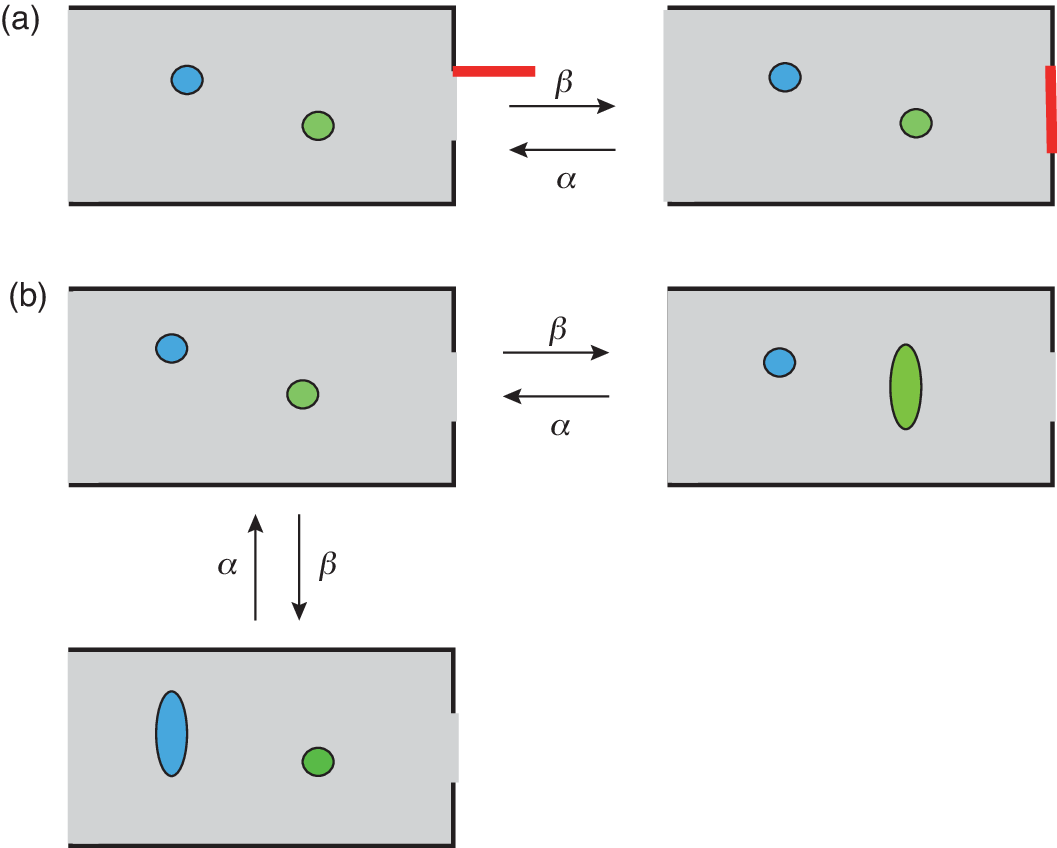}
\caption{Schematic diagram comparing global versus local gating. (a) Global gating: Particles diffuse in a common switching environment consisting of a stochastically gated boundary at $x=L$. (b) Local gating: each particle independently switches between two configurations, only one of which can exit the boundary at $x=L$. (2D representation is for illustrative purposes.)}\label{fig4}
\end{figure}

\begin{figure}[h!]
\centering
\includegraphics[width=13cm]{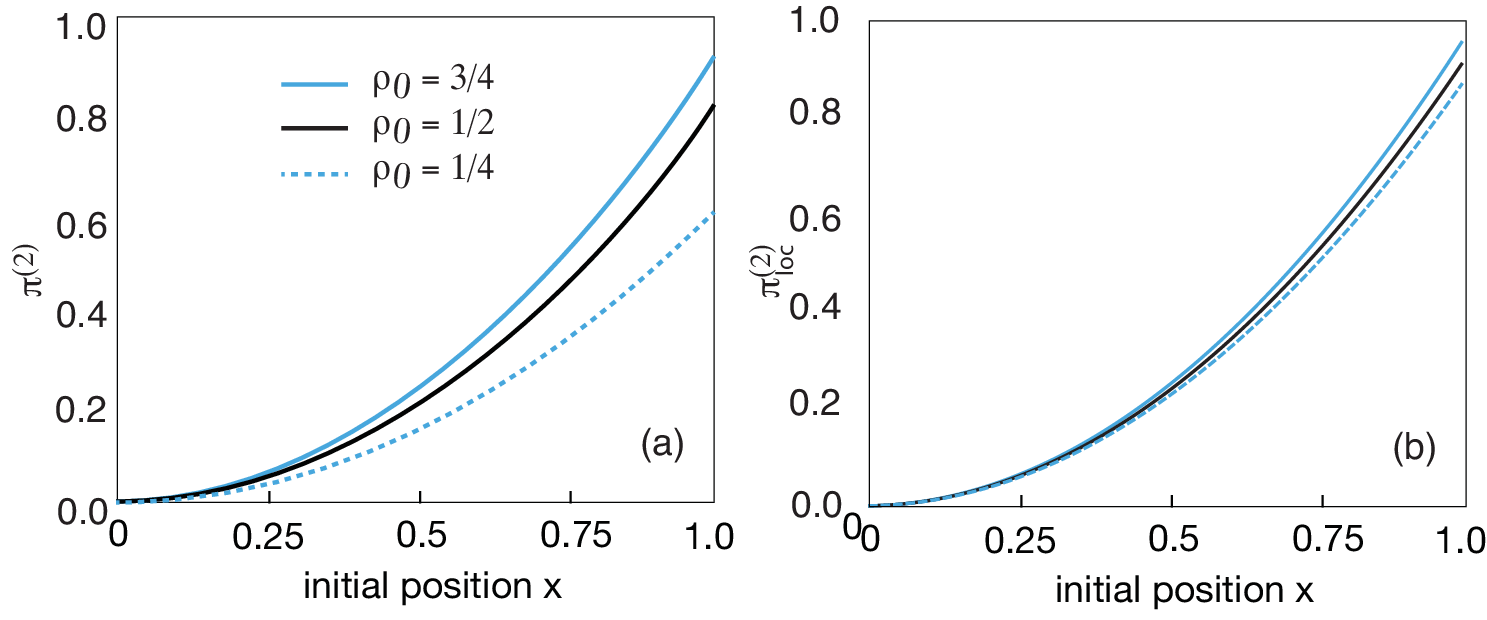}
\caption{(a) Plot of the splitting probability $\pi^{(2)}(x,x)$ for a pair of Brownian particles with the same initial position $x$. The Fourier series solution (\ref{genC}) is truncated after 200 terms. In addition, a 50,000-dimensional version of the infinite-dimensional matrix equation~(\ref{hat}) is solved to estimate the Fourier coefficients. Other parameters are $L=1$ and $\xi=\sqrt{(\alpha+\beta)/D}=10$. (b) Corresponding plot of $\pi_{\rm loc}^{(2)}(x,x)\equiv \pi(x)^2$, which is the separable splitting probability for local gating.}\label{fig5}
\end{figure}

We now compare the joint splitting probability $\pi^{(2)}(x,y)$ for a pair of particles with global gating to the corresponding quantity with local gating. In the latter case, we assume that each particle independently switches between two configurations,  only one of which can exit the boundary at $x=L$. This contrasts with global gating in which the boundary itself switches between an open and closed state, see Fig. \ref{fig4}. It follows that
\begin{equation}
 \pi_{\rm loc}^{(2)}(x,y) =\pi(x)\pi(y),
\end{equation}
with $\pi(x)$ given by equations (\ref{gpi}).
For the sake of illustration, we take the initial/reset position of both particles to be the same,  that is, we set $y=x$. In Fig. \ref{fig5}(a) we plot the splitting probability $\pi^{(2)}(x,x)$ for global gating as a function of $x$ and different choices of $\rho_0$. The corresponding plots of $\pi^{(2)}_{\rm loc}(x,x)$ are shown in Fig. \ref{fig5}(b). It can be seen that in the given parameter regime, correlations due to global gating reduce the splitting probability. 

\bigskip

\section*{References}

  \bibliographystyle{iopart-num}

\begin{thebibliography}{10}


\bibitem{Biroli23} Biroli M, Larralde H, Majumdar S N and  Schehr G 2023 Extreme statistics and spacing sistribution in a Brownian gas correlated by resetting. {\em Phys.
Rev. Lett. } {\bf 130} 207101.

\bibitem{Biroli24b} Biroli M, Kulkarni M, Majumdar S N and  Schehr G 2024 Dynamically emergent correlations between particles in a switching harmonic trap. {\em Phys. Rev. E} {\bf  109} L032106 (2024).

 \bibitem{Biroli26a}  Biroli M, Majumdar S N and  Schehr G 2026 First-passage resetting gas. {\em EPL} {\bf 153} 31002.

\bibitem{Biroli26b} Biroli M, Ciliberto S, Kulkarni M, Majumdar S N, Petrosyan A and  Schehr G 2025 Experimental evidence for strong emergent correlations between particles in a switching trap. Preprint. arXiv:2508.07199


\bibitem{Boyer26}
 Boyer D and Majumdar S N 2025 Emerging correlations between diffusing particles evolving via simultaneous resetting with memory. arXiv:2510.21972.
 


\bibitem{Bressloff15a} Bressloff P C and Lawley S D 2015 Moment equations for a piecewise deterministic PDE. {\em J. Phys. A} {\bf 48} 105001 

\bibitem{Bressloff15b} Bressloff P C and Lawley S D 2015 
Escape from subcellular domains with randomly switching boundaries. {\em Multiscale Model. Simul.} {\bf 13} 1420-1455  

\bibitem{Bressloff16} Bressloff P C 2016 Diffusion in cells with stochastically gated gap junctions. {\em SIAM J. Appl. Math.} {\bf 76} 1658-1682  

\bibitem{Bressloff16a} Bressloff P C 2016  Stochastic Fokker-Planck equation in random environments. Phys. Rev. E 94 042129 (2016).

\bibitem{Bressloff24a} Bressloff P C 2024 Global density equations for interacting particle systems with stochastic resetting: from overdamped Brownian motion to phase synchronization. {\em Chaos} {\bf 34}, 043101 

\bibitem{Bressloff25} Bressloff P C 2025 Stochastic calculus of run-and-tumble motion: an applied perspective. {\em Proc. Roy Soc. A} {\bf 481} 20240815.
 \bibitem{Evans11a} Evans M R and Majumdar S N 2011 Diffusion with stochastic resetting {\em Phys. Rev. Lett.}{\bf 106} 160601.
 
 \bibitem{Mauro26}
de Mauro G,  Biroli M, Majumdar S N and  Schehr G 2026 Dynamically emergent correlations in Brownian particles subject to simultaneous non-Poissonian resetting protocols. {\em Phys.
Rev. E} {\bf 113} 014120.



\bibitem{Evans11b} Evans M R and Majumdar S N 2011 Diffusion with optimal resetting {\em J. Phys. A Math. Theor.} {\bf 44} 435001. 


\bibitem{Evans14} Evans M R and Majumdar S N 2014 Diffusion with resetting in arbitrary spatial dimension {\em J. Phys. A: Math. Theor.} {\bf 47} 285001



\bibitem{Evans20} Evans M R and Majumdar S N, Schehr G 2020 Stochastic resetting and applications {\em J. Phys. A: Math. Theor.} {\bf 53} 193001.


\bibitem{Lawley16} Lawley S D 2016 Boundary value problems for statistics of diffusion in a randomly switching environment: PDE and SDE perspectives
{\em SIAM Journal on Applied Dynamical Systems} {\bf  15}, 1410-1433

\bibitem{Magdziarz22} Magdziarz M and T\'azbierski K 2022 Stochastic representation of processes with resetting. {\em Phys. Rev. E} {\bf 106}, 014147  

\bibitem{Majumdar26} Majumdar S N and Scher G 2026 Dynamically emergent correlations. Preprint. arXiv:2603.03162v1



\bibitem{Mesquita25a}  Mesquita N, Majumdar S N and Sabhapandit S 2025 Dynamically emergent correlations in a Brownian
gas with diffusing diffusivity {\em J. Stat.
Mech.} {\bf 103207}


\bibitem{Pal19} Pal A and Prasad V V 2019 First passage under stochastic
resetting in an interval {\em Phys. Rev. E} {\bf 99} 032123

\bibitem{Sabha24} Sabhapandit S and Majumdar S N 2024 Noninteracting particles in a harmonic trap with a stochastically driven center. {\em J. Phys. A: Math. Theor.}
{\bf 57} 335003.

\end{thebibliography}

\end{document}